\documentclass[oldversion, longauth]{aa}
\usepackage{txfonts}
\usepackage{graphicx}

\begin{document}

\title{Discovery of VHE gamma-ray emission from 1ES\,1218+30.4}

\author{
 J.~Albert\inst{1} \and 
 E.~Aliu\inst{2} \and 
 H.~Anderhub\inst{3} \and 
 P.~Antoranz\inst{4} \and 
 A.~Armada\inst{2} \and 
 M.~Asensio\inst{4} \and 
 C.~Baixeras\inst{5} \and 
 J.~A.~Barrio\inst{4} \and 
 M.~Bartelt\inst{6} \and 
 H.~Bartko\inst{7}
% ...
}
 \institute {Universit\"at W\"urzburg, D-97074 W\"urzburg, Germany
 \and Institut de F\'\i sica d'Altes Energies, Edifici Cn., E-08193 Bellaterra (Barcelona), Spain
 \and ETH Z\"urich, CH-8093 H\"onggerberg, Switzerland
 \and Universidad Complutense, E-28040 Madrid, Spain
 \and Universitat Aut\`onoma de Barcelona, E-08193 Bellaterra, Spain
 \and Universit\"at Dortmund, D-44227 Dortmund, Germany
 \and Max-Planck-Institut f\"ur Physik, D-80805 M\"unchen, Germany
% ...
}

\date{Received date / Accepted date}

\abstract
{
The MAGIC collaboration has studied the high peaked BL-Lac object
1ES\,1218+30.4  at a redshift $z$\,$=$\,$0.182$,  using the MAGIC
imaging air Cherenkov telescope located on the Canary island of La
Palma. A gamma-ray signal was observed with 6.4\,$\sigma$ significance.
The differential energy spectrum for an energy threshold of 120\,GeV 
can be fitted by a simple power law yielding 
$F_E(E) = (8.1\pm 2.1) \cdot 10^{-7} (E/\rm{250\,GeV})^{-3.0\pm
0.4}\,\rm{TeV}^{-1}\,\rm{m}^{-2}\,\rm{s}^{-1}$. During the six days of
observation in January 2005 no time variability on time scales of days
was found within the statistical errors. The observed integral flux
above 350\,GeV is nearly a factor two below the the upper limit
reported by the Whipple Collaboration in 2003.}


\titlerunning{Discovery of VHE gamma-ray emission from 1ES\,1218+30.4}
\authorrunning{J.~Albert et al.}

\maketitle

\section{Introduction}

Since the detection of blazars as sources of high energy
gamma-rays by EGRET on board the CGRO,
the search for the very high energy (VHE) counterparts of these sources 
has been a major goal for ground-based gamma-ray astronomy.  
The first blazar detected at VHE energies was Mrk\,421
(\cite{pun}). 
Subsequent to the discovery of Mrk\,421, only a small number of
extragalactic VHE sources was found, although the sensitivity of
ground-based telescopes is superior in their energy range to that of
EGRET. The total number of VHE blazars reported in the literature as of
February 2006 amounts to 11,
while the number of high-confidence blazars in the 3rd EGRET catalogue
is 66 (\cite{har}).

Stecker et al.\ \cite{ste} pointed out that the
attenuation of gamma-rays due to photon-photon interactions with low
energy photons from the extragalactic background radiation was a likely
explanation of this deficit.  In fact, the redshifts of blazars
detected so far above 100\,GeV are at rather low values 
as expected from predictions of the correlation between the
gamma-ray attenuation and the redshift of the source, known as
Fazio-Stecker relation (\cite{fazio}; \cite{kneiske}). 
These primary gamma-rays, absorbed by pair production, turn into
secondary gamma-rays
contributing to the extragalactic background light (\cite{ahacoppi}).

The VHE gamma-ray emitting blazars known so far all belong to the
class of high-frequency-peaked BL Lacertae
objects (HBLs, \cite{HBL}), a subclass of blazars characterised by
a low luminosity when compared with quasars and a synchrotron peak in
the X-ray band. Their Spectral Energy Distribution (SED) is characterized
by a second peak at very high gamma-ray energies. In
synchrotron-self-Compton (SSC) models, it is assumed that the observed
gamma-ray peak is due to the inverse-Compton emission
from the accelerated electrons
up-scattering previously produced synchrotron photons to high
energies 
(\cite{mar}).
In hadronic models, secondary electrons and positrons,
arising from photo-meson production, initiate electromagnetic cascades 
(\cite{man}; \cite{pro}).
A compilation of blazars with known X-ray
spectra allowing their classification as HBLs is given by \cite{don}.

During the first year of operation
of the Major Atmospheric Gamma-ray Imaging Cherenkov (MAGIC) telescope,
a sample of X-ray bright ($F_{1\,\rm{keV}}$\,$>$\,$2\,\rm{\mu Jy}$),
northern HBLs at moderate  redshifts ($z$\,$<$\,$0.3$), including
1ES\,1218+30.4, 
which is located at a redshift $z$\,$=$\,$0.182$ (\cite{ver}) 
have been observed.
Predictions based on models invoking a synchrotron
self-Compton (\cite{CostGhis}) or hadronic (\cite{man2}) origin of
the gamma-rays and the shape of the time-averaged SED 
show that these HBLs should be detectable by imaging air Cherenkov
telescopes (IACTs), although most of them are too faint
in the EGRET energy range to were detected during its all-sky
survey.

For 1ES\,1218+30.4, the EGRET upper limit is
$F($\,$>$\,$100\,\mbox{MeV})$\,$\approx$\,$10^{-11}\,
\mbox{erg}\,\mbox{cm}^{-2}\,\mbox{s}^{-1}$. 
Several observations with the Whipple telescope between 1995 and 2000
resulted in an upper flux limit of $F(>350\,\mbox{GeV})=8.3 \cdot 10^{-8}\,
\mbox{photons}\,\mbox{m}^{-2}\,\mbox{s}^{-1}$ (corresponding to
$\sim$8\% of the Crab Nebula flux) (\cite{Whipple_AGN}). The source was
also observed 
by HEGRA between 1996 and 2002. An upper flux limit above 840\,GeV of $2.67
\cdot 10^{-8}\,\mbox{photons}\,\mbox{m}^{-2}\,\mbox{s}^{-1}$ (or 12\%
of the Crab Nebula flux) is reported by \cite{ahaakh}.

In this letter, we report the first detection of VHE gamma-rays
($>$\,$100\,\mbox{GeV}$) from the direction of 1ES\,1218+30.4. In Sect.\,2 we
briefly discuss the observations and the data set while we describe in
Sect.\,3 the data analysis method and the results. In Sect.\,4 we
discuss the results in the context of the spectral energy distribution
of 1ES\,1218+30.4 across a broad energy range.

\section{Observations}
The MAGIC telescope is a single dish IACT, located on the Canary island
of La Palma (28.8\degr\,N, 17.8\degr\,W, 2200\,m a.s.l.). A 17\,m
diameter tessellated parabolic mirror with a total surface of
$234\,\mbox{m}^2$, mounted on a light-weight space frame made from
carbon fiber reinforced plastic tubes, focuses Cherenkov light
from air showers, initiated by gamma-rays or charged cosmic rays,
onto a 576-pixel photomultiplier (PMT) camera with a field-of-view of
$3.5\degr$. 
The analogue signals are transported via optical fibers to the trigger
electronics and each channel is read out by a 300\,MHz FADC. Further
details on the telescope can be found in \cite{MAGIC-commissioning} and
\cite{CortinaICRC}.

MAGIC observed 1ES\,1218+30.4 from January 9th to 15th 2005 at zenith
angles (ZA) between $1.5\degr$ and $13\degr$
during six moonless nights, for a total observation time of 8.2\,h. 
To determine the background, so-called
off-data were taken in addition, by pointing the telescope to a nearby
sky region where no gamma-ray source was expected. The off-data cover
the same ZA range with a similar night-sky background light intensity.
For the analysis, 6.5\,h off-data, taken between January 9th and
January 11th, were used, which match the observation conditions and
detector performance of the on-data.
At the same time the KVA telescope
(\textit{http://tur3.tur.iac.es/}) on La Palma
observed 1ES\,1218+30.4 in the optical range.

The signal was discovered with our automatic analysis
(\cite{dor}). The final
analysis is decribed hereafter.

\section{Data analysis and results}
The data were processed using the MAGIC
Analysis and Reconstruction Software (MARS) (\cite{mars}). A description
of the different analysis steps can be found in \cite{gaug} (including
the calibration) and \cite{bre2}.
The moments up to third order of the light distribution are used to
characterise each event by a set of image parameters
(\cite{Hillas_parameters}).

For background suppression, a SIZE-dependent parabolic cut in
WIDTH $\times$ LENGTH is applied (\cite{dyn_cuts}). 
\begin{figure}[hb]
\resizebox{\hsize}{!}{\includegraphics{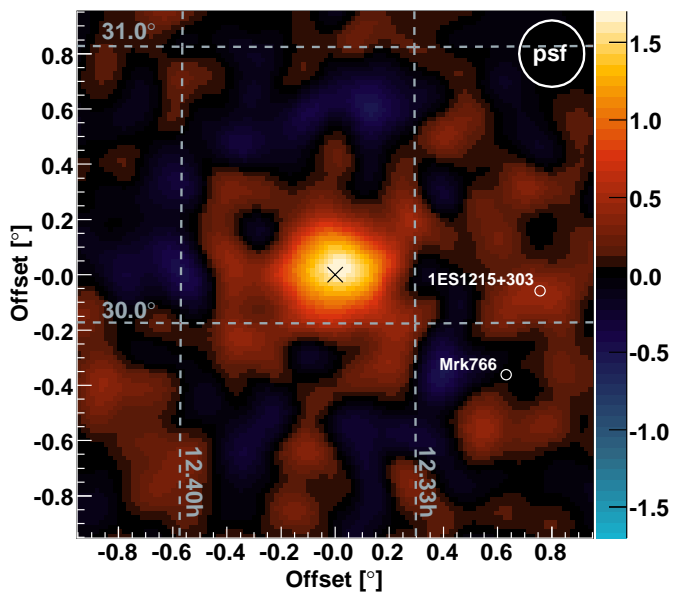}} 
\caption{Sky map of the region around the position of
1ES\,1218+30.4 (cross) for gamma events with an energy threshold of
$\sim$140\,GeV. 
The scale is in units of 10\,events/$\mbox{arcmin}^2$.}
\label{fs}
\end{figure}

To reconstruct the origin of the shower in the camera plane, the DISP
method is employed (\cite{les}) to estimate
the distance between the centre of gravity of the shower and its
origin. The third moment determines the direction of the shower
development. The constant coefficient $\xi$ from the parameterisation
of DISP in the original approach is replaced by $\xi_0 +
\xi_1\cdot \left(\mbox{LEAKAGE}\right)^{\xi_2}$,
LEAKAGE being the fraction of light contained in the outermost camera
pixels. Thereby the truncation of shower images at the camera border is
taken into account. These coefficients were determined using simulated
data.
Simulated gamma-showers (MC) were produced by 
CORSIKA, version 6.023 
(\cite{cor}; \cite{maj}) for ZA below $20\degr$ and
energies between 10\,GeV and 30\,TeV, following a power law with a
spectral index -2.6.
The cut coefficients for the background suppression were
optimised using data of the Crab Nebula, taken at similar ZA in a time
slot of several weeks around the observation.

Figure\,\ref{fs} shows the background-subtracted distribution of
reconstructed shower origins centred at the position of 1ES\,1218+30.4.
Each shower origin is folded with a two-dimensional Gaussian with a
standard deviation of $0.06\degr$, corresponding to
half of the sigma of the point-spread function expected for a
point-like gamma-ray source as seen by MAGIC. The centre of gravity of
the observed excess coincides within the systematic pointing
uncertainty of $0.04\degr$ with the nominal source position.

\begin{figure}[h]
\resizebox{\hsize}{!}{\includegraphics{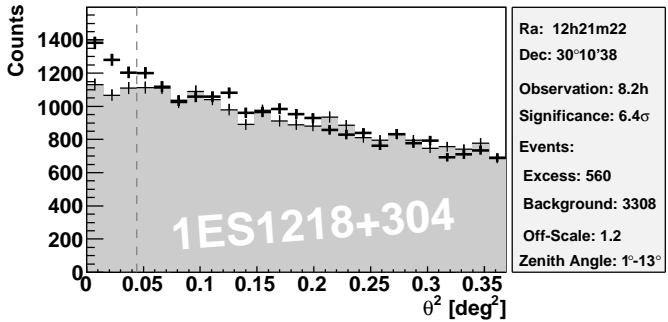}}
\caption{Distribution of $\theta^2$ for on- and off-data
(grey shaded). The signal region is marked by the dashed line.} 
\label{theta}
\end{figure}

Figure\,\ref{theta} shows the distribution of the squared angular
distance, $\theta^2$, between the reconstructed and the nominal source
position, for both the on- and off-data. The background distribution is
not flat due to  the limited trigger area, which means that the highest
acceptance of the detector is in the center and decreases outwards. Due
to different observation times, the off-data were scaled to match the
on-data in the region $0.14<(\theta/$\degr$)^2<0.64$, where no bias
from the source is expected. The observed excess of 560 events has a
statistical significance of 6.4 standard deviations (according to
\cite{lima}). The energy threshold, defined as the peak energy of
simulated gamma-rays with a differential spectrum $\propto$\,$E^{-3.0}$
surviving all cuts, is $\sim$140\,GeV. 
\begin{figure}
\resizebox{\hsize}{!}{\includegraphics{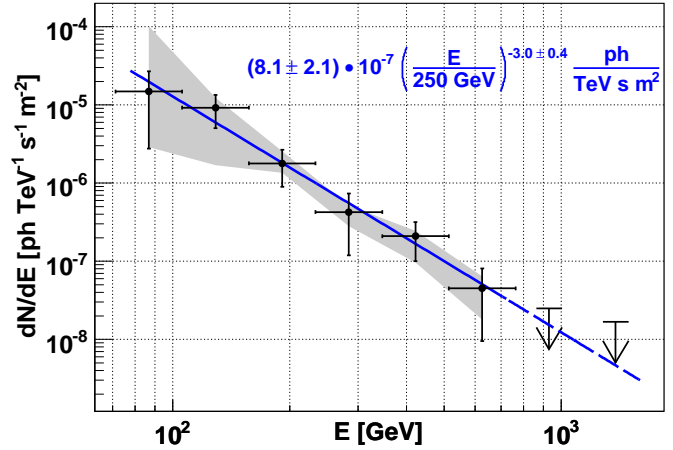}}
\caption{Differential energy spectrum of 1ES\,1218+30.4. The upper limits
correspond to a 90\% confidence level.}
\label{spec}
\end{figure}

For the energy estimation, the Random Forest regression method was applied
(\cite{brei}), trained by the above mentioned MC sample, yielding an
average energy resolution of 24\%. 
Figure\,\ref{spec} and table\,\ref{data} show the reconstructed energy
spectrum. 
To avoid systematic effects of cut-efficiencies to the MC-sample, the
acceptance window for gamma-rays is enlarged, yielding an energy
threshold of $\sim$120\,GeV.
Spill-over effects to neighbouring bins 
were corrected by the number ratio between the energy distribution of a
MC sample compared to the distribution of estimated energy.
A power law is fitted to the measured spectral points yielding
($\chi^2 /NDF = 1.1/4$):
\begin{equation}
\label{diffspec}
F_E(E) = (8.1\pm 2.1) \cdot 10^{-7}\,\left(\frac{E}{\mbox{250\,GeV}}\right)^{-3.0\pm 0.4}\,
\mbox{m}^{-2}\,\mbox{s}^{-1}\,\mbox{TeV}^{-1}
\end{equation}

The error bars ($1\sigma$) show the statistical uncertainty only.
The systematic error resulting from the analysis was investigated
by changing cut-coefficients and the initial MC spectrum (grey-shaded
region).
The total systematic error for the spectral slope is estimated to be
+\,0.7 -\,0.4. Additional systematics for the flux level are estimated to
be in the order of $\sim$40\%. The main contributions are uncertainties
of the atmospheric conditions, the mirror reflectivity and the
effective quantum efficiency of the PMTs.
\begin{table}[h]
\caption{Differential flux of 1ES\,1218+30.4 and the number of measured
excess events together with the statistical error. For the flux also
the systematic error is given (see text for details).}
\label{data}
\centering
\begin{tabular}{c c c} 
\hline \hline 
  E [GeV] & $F_E(E) \pm \mbox{stat +sys -sys}$
[$\mbox{m}^{-2}\,\mbox{s}^{-1}\,\mbox{TeV}^{-1}$] &
$\mbox{N}_{\mbox{exc}} \pm \mbox{stat}$\\
\hline 
\ \ 87  & $(1.5 \pm 1.2 +8.7 -1.2) \cdot 10^{-5}$ & $27 \pm 21$\\
130 & $(9.2 \pm 4.2 +3.1 -7.5) \cdot 10^{-6}$ & $163 \pm 73$\\
190 & $(1.8 \pm 0.9 +0.8 -0.4) \cdot 10^{-6}$ & $146 \pm 71$\\
280 & $(4.3 \pm 3.1 +0.9 -1.4) \cdot 10^{-7}$ & $64 \pm 46$\\
420 & $(2.1 \pm 1.1 +0.4 -1.1) \cdot 10^{-7}$ & $61 \pm 30$\\
630 & $(4.5 \pm 3.5 +1.9 -2.7) \cdot 10^{-8}$ & $22 \pm 17$ \\
\hline
\end{tabular}
\end{table}

The data were also analysed with two other, independent
analysis techniques, using dynamical super-cuts (\cite{kra}) or the
Random Forest method for background suppression, as used in previous
MAGIC observations (e.g.\ \cite{alb}).
Within the statistical errors, the same significance,
flux and differential spectrum were obtained.

\begin{figure}[ht]
\resizebox{\hsize}{!}{\includegraphics{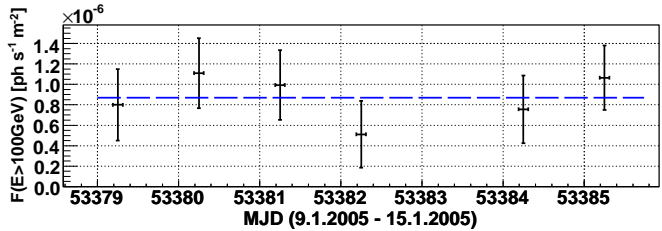}} 
\caption{Integral flux (\,$>$\,$100\,\mbox{GeV}$) of
1ES\,1218+30.4 for every sub-sample. The error bars
($1\sigma$) show the statistical uncertainty only.
}
\label{light}
\end{figure}

For the analysis of time variability, the sample was divided into six
sub-samples, one for every night of observation, with $\sim$85\,min
observation time each. Figure\,\ref{light} shows the integral flux
above 100\,GeV for each night. 
Within the statistical errors, we get a constant
flux ($\chi^2 /NDF = 2.4/5$) of
$F($\,$>$\,$100\,\mbox{GeV})=(8.7\pm1.4) \cdot
10^{-7}\,\mbox{photons}\,\mbox{m}^{-2}\,\mbox{s}^{-1}\,.
$
The
systematic error is in the order of $\sim$40\% as discussed above.
The integral flux is $\sim$30\% higher than one would expect from
integration of Eq.\,(\ref{diffspec}). This can be understood, taking
into account that the flux at 130\,GeV (peak of the event distribution)
lies $\sim$35\% above the fit value. 

\section{Discussion and conclusions}

The discovery of VHE gamma-ray emission from 1ES\,1218+30.4 provides
further evidence that HBLs generally exhibit gamma-ray emission at a
luminosity comparable to the X-ray luminosity.
Archival X-ray data show that the X-ray flux at 1\,keV varies in the
range 0.8 to 4.5  in units of 
$10^{-11}\,\mbox{erg}\,\mbox{cm}^{-2}\,\mbox{s}^{-1}$  (\cite{ros}),
which is below the sensitivity of the All Sky Monitor (ASM) on board
the RXTE X-ray satellite. The observed gamma-ray flux at 100\,GeV is
with $(1.9$\,$\pm$\,$1.1)$\,$\cdot$\,$
10^{-11}\,\mbox{erg}\,\mbox{cm}^{-2}\,\mbox{s}^{-1}$ at the same level
as published X-ray fluxes. Contemporaneous observations from Swift
in October 2005 showed a flux comparable to that measured by BeppoSAX
in 1999 (\cite{gio}), which is also within the range of the one
measured by ROSAT.

There are no indications from simultaneous optical observations (KVA)
of a flare to have occurred during the time of the gamma-ray
observations, neither does the gamma-ray light-curve show signs of
significant variability.
The spectrum is consistent with upper limits at higher
energies determined in the past.

Taken the upper limit of EGRET into account, the gamma-ray peak of the
SED is constrained to lie in the 1\,GeV to 100\,GeV regime, unless
attenuation of the gamma-rays by pair
production in the metagalactic radiation field masks an emitted bump in
the spectrum at higher energies. Detailed modelling of the attenuation
thus becomes the task to unfold the emitted spectrum from the observed
one. Sizeable attenuation is expected from current models of the evolving
extragalactic (=metagalactic) background light (\cite{dwek};
\cite{kneiske}).

At hard X-ray energies, SSC models predict a sharp roll-over in the
SED, in contrast to the predictions of hadronic models involving
electromagnetic cascades. Observations in the hard X-ray domain, e.g.\
using Swift or Suzaku (Astro-E2), and in the intermediate gamma-ray
regime with GLAST, will be crucial to provide information about the
missing link between the synchrotron peak and the gamma-ray peak.

\begin{acknowledgements}
We would like to thank the IAC for the excellent working
conditions at the Observatorio del Roque de los Muchachos in La Palma.
The support of the German BMBF and MPG, the Italian INFN and the
Spanish CICYT is gratefully acknowledged. This work was also
supported by ETH Research Grant TH~34/04~3 and the Polish MNiI Grant
1P03D01028.
\end{acknowledgements}

\end{document}